\documentclass[conference,letterpaper]{IEEEtran}

\usepackage[utf8]{inputenc} 
\usepackage[T1]{fontenc}
\usepackage{url}
\usepackage{ifthen}
\usepackage{graphicx}
\usepackage[subfigure]{tocloft}
\usepackage{subfigure}
\usepackage{cite}
\usepackage[cmex10]{amsmath} 
\usepackage{lipsum}
\newtheorem{theorem}{Theorem}

\newtheorem{lemma}{Lemma}

\newtheorem{definition}{Definition}
\newtheorem{proof}{Proof}

\begin{document}
\title{Theoretical Limit of Radar Time Delay Estimation} 


\author{%
  \IEEEauthorblockN{Dazhuan Xu, Han Zhang, Weilin Tu, Chao Shi, Ying Zhou}
  \IEEEauthorblockA{Nanjing University of Aeronautics and Astronautics\\
                    College of Electronic and Information Engineering, Nanjing  China\\
                    Email:{\{xudazhuan, zhanghan93, tuweiling, shichao, zhouying\}@nuaa.edu.cn}}
}


\maketitle

\begin{abstract}
  In this paper, we proposed the theoretical limit of radar ranging and a limit-achieving time delay estimation method. Based on the prior distribution of traget’s time delay and scattering properties of tragets, posterior probability distribution of traget’s range is derived. Entropy error is defined as power of posterior differential entropy, which can be used to evaluate the performance of time delay estimation methods. Entropy error bound is then put forward, which is independent of estimation methods. 
  A stochastic parameter estimation method is proposed by sampling a posteriori probability distribution. With the definition of the jointly typical sequence and the Chebyshev inequality, we prove that the entropy error of sampling a posteriori probability can approach the entropy error bound when the snapshot number is tending to infinity. 

\end{abstract}

\section{INTRODUCTION}
The typical goal of radar is to detect, localize, and track targets based on the reflected echoes \cite{1,2}. The echoes can be exploited to extract useful information of the targets \cite{3,4}, including range, velocity, shape, and angular direction. The investigation on quantitative problem of extracted information dates back to the 1950s. Woodward and Davies \cite{5} adopted the inverse probability principle to study the mutual information and obtained the approximate relationship among the range mutual information, the time-bandwidth product and the SNR of a single target with constant coefficient \cite{6}. With the seminal work by Bell in 1988 \cite{7}, mutual information regained its footing in radar signal processing to adaptively design the transmitting waveform, which can extract more target-information from the received measurements \cite{8,9}. Surprisingly, there is is very limited literature that focus on the quantitative problem since then for more than seventy years.

In this paper, we introduce a radar system model \cite{10}, which is equivalent to a communication system with joint amplitude, phase, and time delay modulation. To determine the desired description of target in space, we define spatial status, which is composed of target’s range and echoed signal. The spatial information is defined as the joint mutual information between spatial status and received signal. Thus, quantitative problems of radar information are solved, radar and communication systems are unified on the basis of Shannon’s information theory.

Based on the prior distribution of time delays and statistical properties of eoched signals, the posterior probability distribution of target’s range is derived. To evaluate time delay estimation methods, a metric called entropy error (EE) is given, which is the entropy power of the posterior probability distribution \cite{11}. Compared to the mean square error (MSE), EE is more universal, for the reason that error are generally not second-order statistics in low and medium SNR region. Generally, the theoretical entropy error refers to the entropy error bound (EEB), which is decided by the spatial status and method-independent. While the empirical EE is the EE of the specific estimation method.

A time delay estimation method is proposed, called sampling a posteriori probability (SAP), which obtains estimation by sampling the posteriori probability distribution. Different from maximum likelihood estimation and maximum a posteriori probability estimation, SAP is a stochastic estimation method, whose performance coincides with the posteriori probability distribution. It is proved that the empirical EE of SAP approaches the EEB when the snapshot number is tending to infinity, on the contrary, the empirical entropy error of any unbiased estimation method is no less than the EEB.

The rest of the manuscript is organized as follows: Section II establishes the model of multi-target detection system. Section III provides the statistical model of target and channel. In Section IV, 
the metric EE and SAP estimation method is proposed. The theoretical bound of radar time delay estimation is proposed and the corresponding proof is provided in Section V. Section VI concludes our work.

\section{Radar parameter estimation System Model}
Suppose there are $\emph{L}$ targets in the observation interval which are independent of each other, and their positions and scattered signals are also independent of each other. Without loss of generality, let $s_l=\alpha_le^{j\phi_l}$ denotes the complex reflection coeffcient of the $\emph{l}$-th target and $d_l$ denote the distance between the $\emph{l}$-th target and the receiver, for $\emph{l}=1,...,\emph{L}$. Down converting the received signal to baseband, we have

\begin{equation}
z(t)=\sum_{l=1}^{L} s_{l} \psi\left(t-\tau_{l}\right)+w(t)
\end{equation}
where $\psi(\cdot)$ denotes the real baseband signal whose bandwidth is $\emph{B}/2$ and carrier frequency is $\emph{f}_c$, then the phase the transmitted signal can be expressed as $\varphi_{l}=-2 \pi f_{c} \tau_{l}+\varphi_{l_0}$, where $\phi_{l_0}$ denotes the initial phase, $\tau_{l}=2 d_{l} / v$ denotes the time delay of the  $l$-th target and $v$ is the signal propagation velocity of the signal. $w(t)$ is the complex additive white gaussian noise (CAWGN) with mean zero and variance $N_0/2$ in its real and imaginary parts respectively. The bandwidth of $w(t)$ is $B/2$.

In general, the amplitude of the scattering coefficient is a function of the time delay and inversely proportional to the distance. For simplicity, it is implicitly assumed that the observation interval is small and the influence of attenuation can be ignored, so the amplitude $\alpha$ is invariant. Although the amplitude of each interval is different, it can be regarded as a constant in each interval. The following analysis method in this paper is still applicable.
 \begin{figure}[htbp]
	\subfigure[Observation distance interval]{\includegraphics[width=0.45\textwidth]{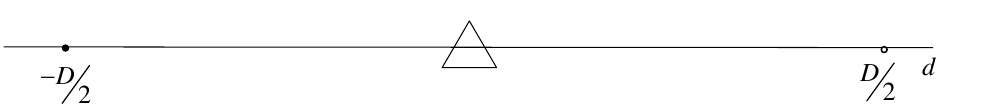}
		\label{f1a}}
	\subfigure[Observation time interval]{\includegraphics[width=0.45\textwidth]{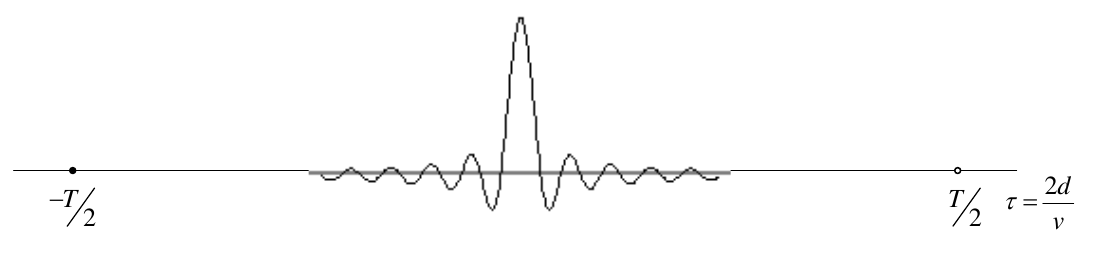}
		\label{f1b}}
	\subfigure[Normalized observation interval]{\includegraphics[width=0.45\textwidth]{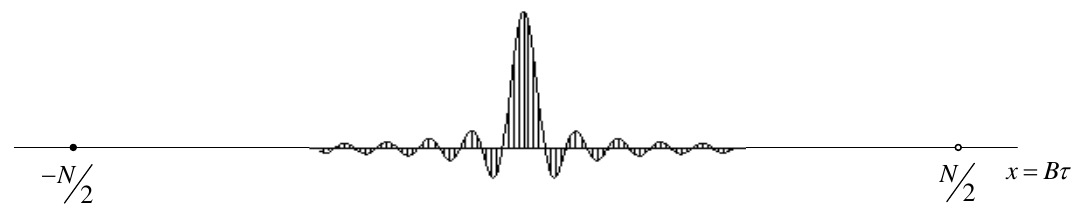}
		\label{f1c}}
	\caption{Three observation intervals and signal waveforms.}
	\label{f1}
\end{figure}

For the convenience of theoretical analysis, it is assumed that the reference point is located at the center of the observation interval and the observation range is $[-D / 2, D / 2)$,  which is shown in Figure \ref{f1a}. The time delay interval is $[-T / 2, T / 2)$, which is shown in Figure \ref{f1b}. It is also assumed that the emitted signal is an ideal low-pass signal and the baseband signal satisfies
\begin{equation}
\psi(t)=\operatorname{sinc}(B t)=\frac{\sin (\pi B t)}{\pi B t},-\frac{T}{2} \leq t \leq \frac{T}{2}
\end{equation}
where $T$ denotes the duration of the signal. 

According to the Shannon-Nyquist sampling theorem, $z(t)$ can be sampled at a rate $B$ to obtain a discrete form of (1)
\begin{equation}
z( {\frac{n}{B}} ) = \sum\limits_{l = 1}^L {{s_l}\psi ( {\frac{{n - B{\tau _l}}}{B}} )}  + w( {\frac{n}{B}} )\;,\;n =  - \frac{N}{2}, \ldots ,\frac{N}{2} - 1
\end{equation}
where, $N=TB$ is the time bandwidth product (TBP). Let $x_l=B\tau_l$ represents the normalization delay of the target, and then the discrete form system equation can be obtained 
\begin{equation}
z\left( n \right) = \sum\limits_{l = 1}^L {{s_l}\psi \left( {n - {x_l}} \right)}  + w\left( n \right)\;,\;n =  - \frac{N}{2}, \ldots ,\frac{N}{2} - 1
\end{equation}
For convenience, express (1) in vector form
\begin{equation}
{\bf{z}} = {\bf{U}}\left( {\bf{x}} \right){\bf{s}} + {\bf{w}}
\end{equation}
where $\mathbf{z}=[z(-N/2), \ldots, z(N/2-1)]^{\mathrm{T}}$ denotes discrete received signal, $\mathbf{s}=[z_1, \ldots, z_L]^{\mathrm{T}}$ denotes target scattering vector, $\mathbf{U(x)}=[u(x_1), \ldots, u(x_L)]^{\mathrm{T}}$  denotes position matrix determined by the transmitted signal waveform and the time delay of target. Its $l$-th column vector $ 
\mathbf{u}( {x_l} ) = [ sin c( { - N/2 - {x_l}} ), \ldots ,z( {N/2 - 1 - {x_l}} ) ]^{\rm{T}}$  is the echo from the $l$-th target with a time delay, ${\bf{w}} = {[ {w( { - N/2} ), \ldots ,w( {N/2 - 1} )}]^{\rm{T}}}$ is the noise vector whose components are independent and identically distributed complex gaussian random variables with mean value of 0 and variance of $N_0$.

\section{Statistical model of target and channel}

Corresponding to that of source in the communication system, the statistical characteristic of target in the radar parameter estimation system is the joint distribution of distance and scattering. 
\begin{equation}
p\left( {{\bf{x,s}}} \right) = p\left( {\bf{x}} \right)p\left( {\bf{s}} \right)
\end{equation}
where $p(\bf{x})$ is the prior probability density function (PDF) of the normalized distance, $p(\bf{s})$ is the PDF of the scattered signal. Generally, target position and scattering are unrelated.

Without prior information, the distance of the target is assumed to obey uniformly distribution in the observation interval.
In this paper, only two typical statistical models of radar electromagnetic scattering signals, constant modulus (Swerling 0) and complex gaussian (Swerling 1), are considered. 

Suppose the channel is CAWGN channel, and the PDF of multi-dimensional complex gaussian noise vector is given by
\begin{equation}p(\mathbf{w})=\frac{1}{\left(\pi N_{0}\right)^{N}} \exp \left(-\frac{1}{N_{0}} \mathbf{w}^{\mathrm{T}} \mathbf{w}\right)\end{equation}
then multi-dimensional PDF of $\bf{z}$ conditioned on $\bf{x}$ and $\bf{s}$ is given by
\begin{equation}p(\mathbf{z} \mid \mathbf{x}, \mathbf{s})=\frac{1}{\left(\pi N_{0}\right)^{N}} \exp \left(-\frac{1}{N_{0}}(\mathbf{z}-\mathbf{U}(\mathbf{x}) \mathbf{s})^{\mathrm{H}}(\mathbf{z}-\mathbf{U}(\mathbf{x}) \mathbf{s})\right)\end{equation}

the above equation defines the channel characteristics of the radar parameter estimation system.

\section{Parameter estimation methods and metric}

\subsection{Maximum likelihood estimation and maximum a posterior probability estimation}

Considering complex gaussian scattering targets with complex additive white Gaussian noise, the received signal $\mathbf{z}$ is also a complex Gaussian vector and its covariance matrix $\mathbf{R}$ is given by 
\begin{equation}\mathbf{R}=E_{\mathrm{S}, \mathrm{W}}\left[\mathbf{z} \mathbf{z}^{\mathrm{H}}\right]\end{equation}
Substituting (5) into (9), we can obtain
\begin{equation}\begin{aligned}
\mathbf{R} &=E\left[(\mathbf{U}(\mathbf{x}) \mathbf{s}+\mathbf{w})(\mathbf{U}(\mathbf{x}) \mathbf{s}+\mathbf{w})^{\mathrm{H}}\right] \\
&=\mathbf{U}(\mathbf{x}) E\left[\mathbf{s} \mathbf{s}^{\mathrm{H}}\right] \mathbf{U}^{\mathrm{H}}(\mathbf{x})+E\left[\mathbf{w} \mathbf{w}^{\mathrm{H}}\right] \\
&=N_{0}\left(\sum_{l=1}^{L} \frac{\rho_{l}^{2}}{2} \mathbf{u}_{l}(\mathbf{x}) \mathbf{u}_{l}^{\mathrm{H}}(\mathbf{x})+\mathbf{I}\right)
\end{aligned}\end{equation}
where $\rho_{l}^{2}=2 E[|s_{l}|^{2}]/N_{0}$ denotes the average SNR for the $l$-th target. Note that the covariance matrix is a function of the range vector $\mathbf{x}$. The probability density function of $\mathbf{z}$ conditioned on $\mathbf{x}$ is given by 
\begin{equation}
p\left( {{\bf{z}}\left| {\bf{x}} \right.} \right){\rm{ = }}\frac{{\rm{1}}}{{{\pi ^N}\left| {\bf{R}} \right|}}\exp \left( { - {{\bf{z}}^{\rm{H}}}{{\bf{R}}^{ - 1}}{\bf{z}}} \right)
\end{equation}
The conditional probability distribution describes the statistical characteristics of the channel, which is also known as likelihood function. The estimation value $\bf{\hat x}$ that maximizes the above equation is called the maximum likelihood estimation of the range $\bf{x}$ and we denote it as ${\bf{\hat x}}_{\rm{ML}}$, then we have
\begin{equation}
{{\bf{\hat x}}_{\rm{ML}}} = \arg \mathop {\max }\limits_{\bf{x}} \frac{{\rm{1}}}{{\left| {\bf{R}} \right|}}\exp \left( { - {{\bf{z}}^{\bf{H}}}{{\bf{R}}^{ - 1}}{\bf{z}}} \right)
\end{equation}

The priori density probability is assumed as $p(\bf{x})$. According to the Bayes formula, the posterior probability distribution is
\begin{equation}p(\mathbf{x} \mid \mathbf{z})=\frac{\frac{1}{|\mathbf{R}|} p(\mathbf{x}) \exp \left(-\mathbf{z}^{\mathrm{H}} \mathbf{R}^{-1} \mathbf{z}\right)}{\oint \frac{1}{|\mathbf{R}|} p(\mathbf{x}) \exp \left(-\mathbf{z}^{\mathrm{H}} \mathbf{R}^{-1} \mathbf{z}\right) d \mathbf{x}}\end{equation}

The estimation value $\bf{\hat x}$ that maximizes the above equation is called the maximum a posterior probability estimation of the range $\bf{x}$ and is denoted as ${{\bf{\hat x}}_{\rm{MAP}}}$
\begin{equation}
{{\bf{\hat x}}_{\rm{MAP}}} = \arg \mathop {\max }\limits_{\bf{x}}p(\mathbf{x} \mid \mathbf{z})
\end{equation}

\subsection{Sampling a posteriori probability estimation}

The estimation value $\bf{\hat x}$ generated by sampling the posterior probability distribution $p(\bf{x}|\bf{z})$ is called sampling a posterior probability ({\rm{SAP}}) estimation of the range $\bf{x}$, which is denoted as $\bf{\hat x}_{\rm{SAP}}$ and is represented as
\begin{equation}
{{\bf{\hat x}}_{\rm{SAP}}} = \arg \mathop{{\rm{sam}}}\limits_{\bf{x}} \frac{{\rm{1}}}{{\left| {\bf{R}} \right|}}p({\bf{x}})\exp \left( { - {{\bf{z}}^{\bf{H}}}{{\bf{R}}^{ - 1}}{\bf{z}}} \right)
\end{equation}

Corresponding to random-coding method, SAP estimation method is a random-estimating method, that is, SAP exhibits no specific rule of estimation. SAP obtains the estimation by sampling the posterior probability in different snapshots. Thus, the performance of SAP estimation entirely depends on the posterior probability distribution, while the performance of other parameter estimation is difficult to determine. 
	
\subsection{Parameter estimation metric}
Assume the posterior probability density of a range estimator is $p(\bf{x}\mid \bf{z})$, the differential entropy of $p(\bf{x}|\bf{z})$ can be represented as 
\begin{equation}
h\left( {x\left| {\bf{z}} \right.} \right) =  - E_{\boldsymbol{z}}\left[-\int_{-T B / 2}^{T B / 2} p(x \mid \boldsymbol{z}) \log p(x \mid \boldsymbol{z}) \mathrm{d} x\right]
\end{equation}

The posterior differential entropy $h(\bf{x}\mid \bf{z})$ represents uncertainty of parameter estimation result. The smaller $h(\bf{x}\mid \bf{z})$ is, the more accurate the estimation result is. Therefore, we propose EE as the metric of range estimator, which is expressed as
\begin{equation}
\sigma _{\rm EE}^2 = \frac{{{2^{2h\left( {\bf{x}\mid \bf{z}} \right)}}}}{{2\pi e}}
\end{equation}
Compared to MSE, EE has a better adaptability to SNR. As error statistics are generally not second-order in medium and low SNR region, it is unreasonable to adopt MSE in these cases.

\section{Entropy error bound of single target}
It is generally considered that the maximum likelihood estimation and the maximum posteriori probability estimation are optimal, which are called the maximum likelihood estimation criterion and the maximum posteriori probability estimation criterion. However, a basic theoretical problem has been neglected for a long time, that is, which estimator is optimal, in what sense is it optimal and what is the optimal performance.

We answer the above three problems in this section. For brevity's sake, only a framework of single-target time delay estimation is considered. Before the proof of parameter estimation theorem, a few definitions are needed.
\subsection{Preparative wroks}
\begin{figure}[htbp]
   \centering
   \includegraphics[width=0.4\textwidth]{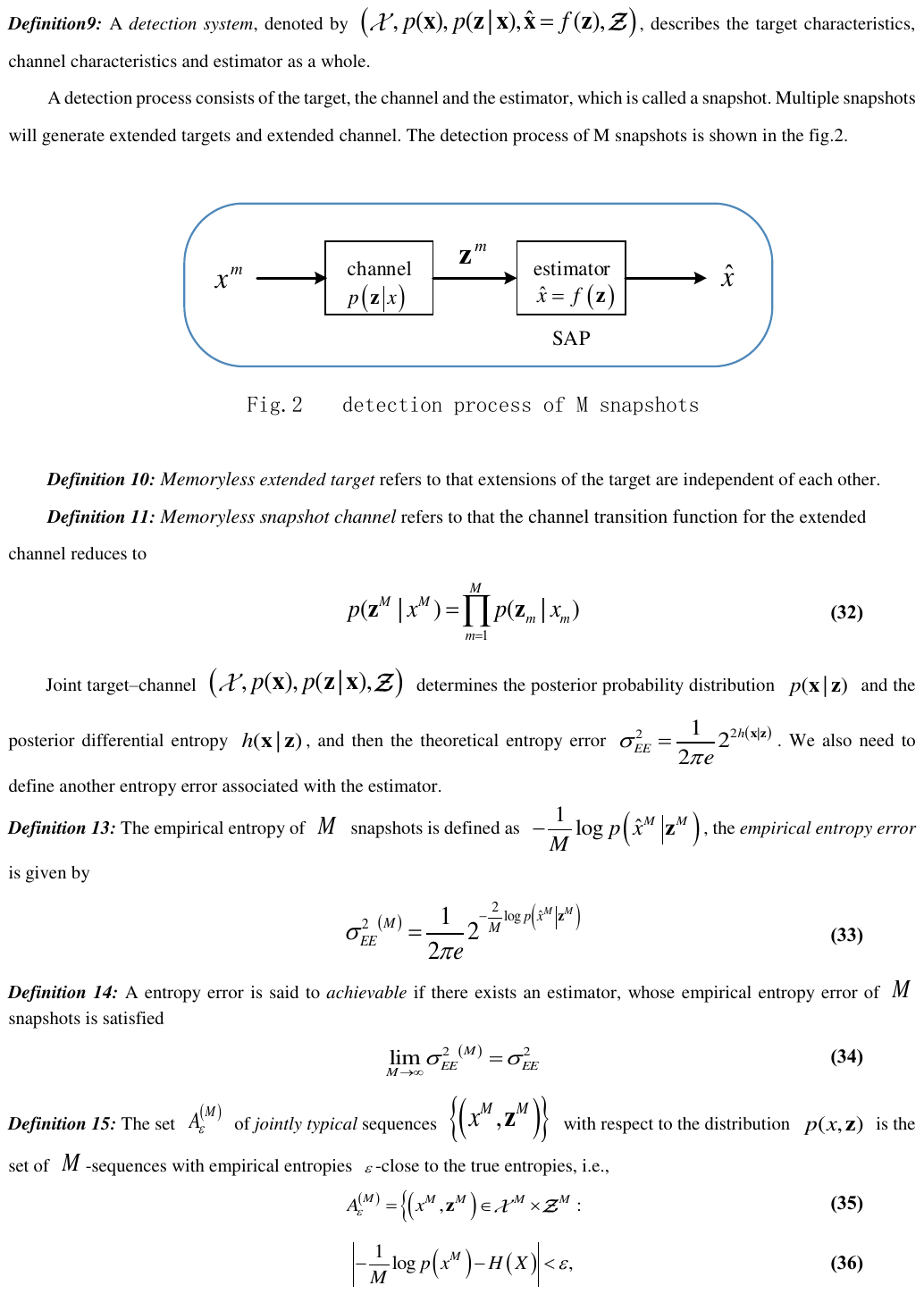}
   \caption{Parameter estimation process of $\emph{M}$ snapshots}
   \label{f2}
 \end{figure}

\begin{definition}
The normalized time delay of a target in the observation interval is a random variable, and the prior distribution of the normalized time delay is called the target delay characteristic or the source statistical characteristic, or simply "target" for short. 
\end{definition}
\begin{definition}
A parameter estimation channel, denoted by $\left( {{\mathcal{X}},p({\bf{z}}|x),{\mathcal Z}} \right)$, consists of two finite sets $\mathcal{X}$, $\mathcal{Z}$ and a collection of probability mass function $p({\bf{z}}|x)$, with the interpretation that the input of the channel  $x$ is the normalized delay of a target defined in the finite real observation interval and the output of the channel $\bf{z}$ is the collection of received complex signal sequence.
\end{definition}
\begin{definition}
An estimator is an estimation function of a normalized delay $\hat{x} = f\left( {\bf{z}} \right)$, which outputs a distance estimation for the given receiving sequence.
\end{definition}
\begin{definition}
A parameter estimation system, denoted by $\left( {{\cal X},p({\bf{x}}),p({\bf{z}}|{x}),{\hat x} = f({\bf{z}}),{\cal Z}} \right)$,  describes the target characteristics, channel characteristics and estimator as a whole. A parameter estimation process consists of the target, the channel and the estimator, which is called a snapshot. Multiple snapshots will generate extended targets and extended channel. The parameter estimation process of $\emph{M}$ snapshots is shown in the figure \ref{f2}.
\end{definition}
\begin{definition}
Memoryless extended target is the extensions of the target are independent of each other. Memoryless snapshot channel is the channel transition function for the extended channel, which satisfies
\begin{equation}
p({{\bf{z}}^M}|{x^M}) = \prod\limits_{m = 1}^M {p({{\bf{z}}_m}|{x_m})}
\end{equation}
\end{definition}
\begin{definition}
Joint target channel $\left( {{\cal X},p({x}),p({\bf{z}}|{x}),{\cal Z}} \right)$ determines the posterior probability distribution $p({x}|{\bf{z}})$ and the posterior differential entropy $h({x}|{\bf{z}})$, the theoretical EEB satisfies
\begin{equation}
\sigma _{EE}^2 = \frac{1}{{2\pi e}}{2^{2h\left( {{x}|{\bf{z}}} \right)}}
\end{equation}

\end{definition}

\begin{definition}
To prove the achievablity of the bound, we define the EE of a estimator as
\begin{equation}  
\sigma {_{EE}^2{^{\left( M \right)}}} = \frac{1}{{2\pi e}}{2^{ - \frac{2}{M}\log p\left( {{{\hat x}^M}\left| {{{\bf{z}}^M}} \right.} \right)}}
\end{equation}
which is called the empirical EE .
\end{definition}
\begin{definition}
	EEB is achievable if there exists an estimator, whose empirical EE of $\emph{M}$ snapshots satisfies
	\begin{equation}
	\lim_{M\rightarrow \infty}{\sigma _{EE}^{2}}^{\left( M \right)}=\sigma _{EE}^{2}
	\end{equation}
\end{definition}
\begin{definition}
The set $A_\varepsilon ^{\left( M \right)}$ of jointly typical sequences $ {\left( {{x^M},{{\bf{z}}^M}} \right)} $ with respect to the distribution $p(x,{\bf{z}})$ is the set of $\emph{M}$ sequences whose empirical entropy differs from the true entropy by less than $\varepsilon $, i.e.,
\begin{equation}
\begin{aligned}
A_{\epsilon}^{M}=&\{(x^{M}, {\bf{z}}^{M}) \in x^{M} \times {\bf{{\bf{z}}}}^{M}:\\&|-\frac{1}{M} \log p(x^{M})-H(x)|<\epsilon,\\&|-\frac{1}{M} \log p({\bf{z}}^{M})-H({\bf{{\bf{z}}}})|<\epsilon,\\
&|-\frac{1}{M} \log p(x^{M}, {\bf{z}}^{M})-H(x, {\bf{z}})|<\epsilon\}
\end{aligned}
\end{equation}
where
\begin{equation}
p\left( {{x^M},{{\bf{z}}^M}} \right) = \prod\limits_{m = 1}^M {p\left( {{x_m},{{\bf{z}}_m}} \right)}
\end{equation}
The input and output of the extended source channel constitute the joint typical sequence.
\end{definition}
\subsection{Theoretical bound and the achievablity}
\begin{lemma}\label{Lemma 1}
	For a memoryless snapshots channel $( {{{\cal X}^M},p({{\bf{z}}^M}|{x^M}),{{\cal Z}^M}})$, if ${\hat{x}}^M$ are $\emph{M}$ sampling estimates of a posterior probability distribution 
	$p(x|{\bf{z}})$, then $(\hat{x}^M,{\bf{z}}^M)$ are jointly typical sequences with respect to the probability distribution $p(\hat{x}^M,{\bf{z}}^M)$.
	\begin{proof}
		Since ${\hat x^M}$ are $\emph{M}$ sampling estimates of a posterior probability distribution 
		$p\left( {x|{\bf{z}}} \right)$, the extended posterior probability distribution ${p_f}\left( {{{\hat x}^M}|{{\bf{z}}^M}} \right) = p\left( {{{\hat x}^M}|{{\bf{z}}^M}} \right)$, then
		\begin{equation}
		\begin{array}{l}
		p_{f}\left(\hat{x}^{M}, \mathbf{z}^{M}\right)=p\left(\mathbf{z}^{M}\right) p_{f}\left(\hat{x}^{M} \mid \mathbf{z}^{M}\right) \\
		=p\left(\mathbf{z}^{M}\right) p\left(\hat{x}^{M} \mid \mathbf{z}^{M}\right)=p\left(\hat{x}^{M}, \mathbf{z}^{M}\right)
		\end{array}
		\end{equation}
		the proof is completed.
	\end{proof}	 
\end{lemma}\label{Lemma 1}

The performance of sampling posteriori probability estimation is completely decided by posterior probability distribution $p\left( {x|{\bf{z}}} \right)$. Therefore, the extended sequence $\left( {{{\hat x}^M},{{\bf{z}}^M}} \right)$ obtained by SAP estimation is jointly typical with respect to probability distribution $p( {{{\hat x}^M},{{\bf{z}}^M}})$.

\begin{theorem}\label{Theorem 2 parameter estimation}
	EEB $\sigma _{EE}^2$ is achieveable. Specifically, given that the estimator knows the joint source-channel statistical characteristics, for any $\varepsilon > 0$, there exists an estimator whose empirical EE satisfies
	\begin{equation}
	\sigma_{EE}^{2} e^{-4 \varepsilon}<\sigma_{EE}^{2(M)}<\sigma_{EE}^{2} e^{4 \varepsilon}
	\end{equation}
	and
	\begin{equation}
	\lim_{M\rightarrow \infty}{\sigma _{EE}^{2}}^{\left( M \right)}=\sigma _{EE}^{2}
	\end{equation}
	Conversely, the empirical EE of any unbiased estimator cannot be smaller than the EEB
\end{theorem}
\begin{proof}
	Consider the following events:

	1. Generate $M$ extensions $x^M$ of the target independently according to distance characteristics of target;

	2. Generate the receiving sequence $z^M$ according to $x^M$ and the characteristics of $M$ extensions of the channel $p({\bf{z}}|x)$, which satisfies
	\begin{equation}
	p({{\bf{z}}^M}|{x^M}) = \prod\limits_{m = 1}^M {p({{\bf{z}}_m}|{x_m})}
	\end{equation}
	
	Introducing SAP estimation method, $\hat{x}^M$ is $M$ sampling estimation, and $( {{{\hat x}^M},{{\bf{z}}^M}} )$ is the jointly typical sequence with respect to probability distribution $p( {{{\hat x}^M},{{\bf{z}}^M}} )$.
	When the snapshot number $M$ is large enough, according to the definition of the jointly typical sequence, for any $\epsilon>0$, 
	\begin{equation}
	\left| { - \frac{1}{M}\log p\left( {{{\hat x}^M},{{\bf{z}}^M}} \right) - H\left( {x,{\bf{z}}} \right)} \right| < \varepsilon
	\end{equation}
	\begin{equation}
	\left| { - \frac{1}{M}\log p\left( {{{\bf{z}}^M}} \right) - H\left( {\bf{z}} \right)} \right| < \varepsilon
	\end{equation}
	as the posterior probability satisfies
	\begin{equation}
	p(\hat{x}^M | {\bf{z}}^M )=p(\hat{x}^M ,{\bf{z}}^M )/p( {\bf{z}}^M )
	\end{equation}
	
	Then we have
	\begin{equation}
	\left| { - \frac{1}{M}\log p\left( {{{\hat x}^M}\left| {{{\bf{z}}^M}} \right.} \right) - H\left( {x\left| {\bf{z}} \right.} \right)} \right| < 2\varepsilon 
	\end{equation}
	
	By the definition of EE and empirical EE
	\begin{equation}
	H\left( {x\left| {\bf{z}} \right.} \right) - 2\varepsilon  <  - \frac{1}{M}\log p\left( {{{\hat x}^M}\left| {{{\bf{z}}^M}} \right.} \right) < H\left( {x\left| {\bf{z}} \right.} \right) + 2\varepsilon
	\end{equation}
	
	Hence
	\begin{equation}
	\sigma_{E E}^{2} e^{-4 \varepsilon}<\sigma_{E E}^{2(M)}<\sigma_{E E}^{2} e^{4 \varepsilon}
	\end{equation}

	Based on the definition of the jointly typical sequence and the Chebyshev inequality, when $\varepsilon  \to 0$ and $M \to \infty $
	\begin{equation}
	\lim _{M \rightarrow \infty} \sigma_{E E}^{2(M)}=\sigma_{E E}^{2}
	\end{equation}
	
	It is proved that the empirical EE of the posterior probability estimator can approach the theoretical EEB.
	
	Converse theorem: The empirical EE of any unbiased estimator is no less than the EEB.
	
	Let ${\hat x^m} = f({{\bf{z}}^m})$ be an arbitrary estimator, and the mutual information obtained by the estimator is ${I_f}\left( {{{\bf{z}}^M},{x^M}} \right)$, then
	\begin{equation}
	I_{f}\left(\mathbf{z}^{M}, x^{M}\right)=h\left(x^{M}\right)-h_{f}\left(x^{M} \mid \mathbf{z}^{M}\right)
	\end{equation}
	where 
	${h_f}( {x^M}| {{{\bf{z}}^M}})$  denotes the posterior differential entropy of the estimator. Obviously,  
	$( {{\bf{z}}^M},{{\hat x}^M},{x^M}) $  forms a Markov chain. Based on data processing theorem,
	\begin{equation}
	{I_f}\left( {{{\bf{z}}^M},{x^M}} \right) \le I\left( {{{\bf{z}}^M},{x^M}} \right)
	\end{equation}
	Then
	\begin{equation}
	{h_f}\left( {{x^M}\left| {{{\bf{z}}^M}} \right.} \right) \ge h\left( {{x^M}} \right) - I\left( {{x^M},{{\bf{z}}^M}} \right) = h\left( {{x^M}\left| {{{\bf{z}}^M}} \right.} \right)
	\end{equation}
	By the definition of EE
	\begin{equation}
	\sigma_{E E}^{2(M)}\left(X^{M} \mid \mathbf{Z}^{M}\right) \geq \sigma_{E E}^{2}
	\end{equation}
	The proof of the Converse theorem is completed.
\end{proof}	 

\subsection{Numerical Examples}
\begin{figure}[htbp]
	\centering
	\includegraphics[width=0.4\textwidth]{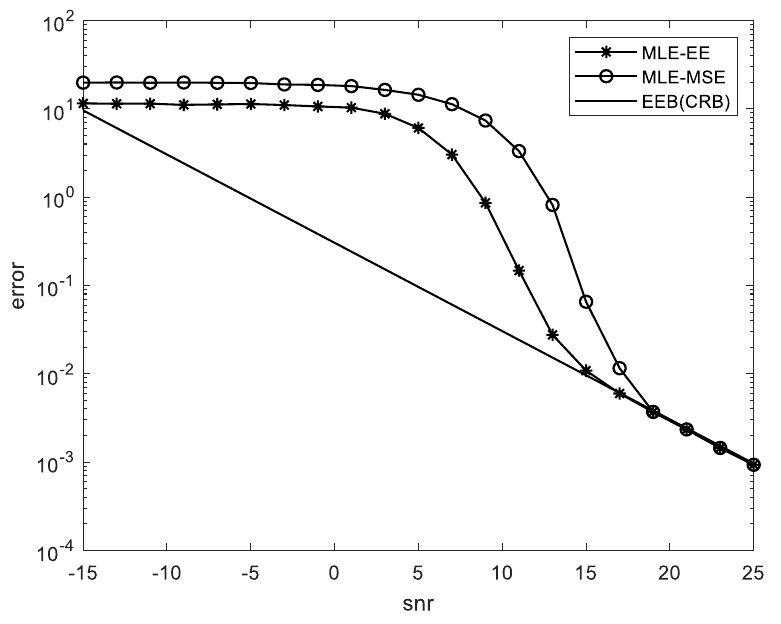}
	\caption{SNR versus error for EE, MSE and EEB with MLE estimation method}
	\label{f3}
\end{figure}

We have established the metric EE and the SAP estimation method. 
It is of interest to see the relationship between EEB and CRB and the performance gap between MLE and SAP in the nonasymptotic regime. To investigate these, we give a numerical study. Corresponding to the proof, only single-target scenario is considered.
As shown in \cite{12}, the approximation of EEB in constant reflection scenario is

\begin{equation}
\sigma _{EE}^2 \ge \frac{{{2^{2{{\log }_2}\frac{{\sqrt {2\pi e} }}{{\rho \beta }}}}}}{{2\pi e}} = {\left( {{\rho ^2}{\beta ^2}} \right)^{ - 1}}
\end{equation}
where $\rho ^2$ denotes signal to noise ratio and $\beta ^2=\pi^2/3$.
In constant amplitudes and complex additive white gaussian noise scenario, set target distance as $x_0=0$ and the observation interval of TBP as $[-8,8]$. Figure \ref{f2} shows
the SNR versus SNR versus error for EE, MSE and EEB with MLE estimation method. Figure \ref{f3} shows
the SNR versus EE tradeoff for MLE estimation, SAP estimation and EEB. 

In figure \ref{f3}, EEB coincides with CRB. So, the solid line denotes EEB and CRB. The line with circle marker is the MSE of MLE estimation, while the line with asterisk marker is the EE of MLE estimation. As can be seen, the line of EE is lower than that of MSE in all SNR region and achieves EEB earlier than MSE. Therefore, EE is more adaptive than MSE in low and medium SNR.

\begin{figure}[htbp]
	\centering
	\includegraphics[width=0.4\textwidth]{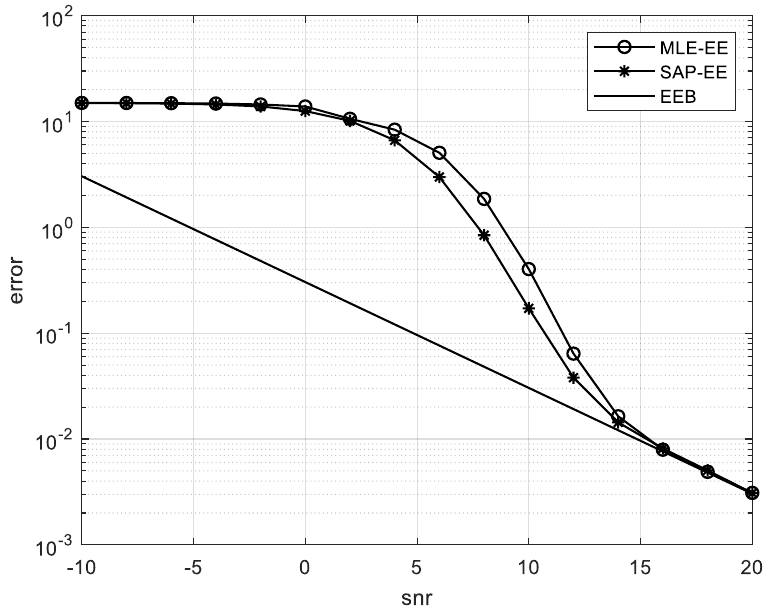}
	\caption{SNR versus EE for MLE estimation, SAP estimation and EEB}
	\label{f4}
\end{figure}

Figure \ref{f4} provides the comparison of the emprical EE of MLE and SAP, the solid line denotes EEB of constant reflection. The line with circle marker is the EE of MLE estimation, while the line with asterisk marker is the EE of SAP estimation. We focus on the nonasymptotic regime. In low SNR regime, the line of SAP is little high than MLE. However, the probability density distribution and posterior distribution do not obeys gaussian distribution. It is more meaningful to pay attention to the medium SNR regime. It can be seen from the figure, SAP has about 1dB performance gain when compared to MLE.
This example shows that SAP estimation shows that SAP estimation approximates the EEB faster than MLE estimation. 

\section{Conclusion}

The mian contribution of this paper are the proposal of EEB and the SAP. EEB is a universal metric and achieveable in low, medium and SNR. EE to parameter estimating is what channel capacity to channel coding and rate-distortion function to source coding, which provides the theoretical bound for all parameter estimation methods.
SAP estimation method is an asymptotically optimal method. The ideology of
SAP estimation is similar to the random coding in Shannon's information theory. Its original
intention is to prove the reachability of the EEB. Also, it is shown that this method is realizable in practical as it avoids the spectral peak search encountered in deterministic estimation method and has the advantage of low complexity in multi-dimensional parameter estimation scenario if relaxation techniques are introduced.    

The closed expression for the bound of EE in specific scenario is an open research subject. Another subject for future work is to explore the theoretical bound of other parameters such as the angular direction.    

\section*{Acknowledgment}

This paper is based upon work supported by the National
Science Foundation under Grant No. 61471192 and No. 61371169.

%



\end{document}